\documentclass[aps,reprint,amsmath,superscriptaddress,amssymb,showpacs]{revtex4-2}
\usepackage{xcolor}
\usepackage{graphicx}
\usepackage{dcolumn}
\usepackage{bm}
\usepackage{amsmath, amssymb, amscd, amsthm, amsfonts}
\usepackage{svg}
\usepackage{subcaption}
\usepackage{hyperref}
\usepackage{physics}

\newcommand{\zstroke}{%
  \text{\ooalign{\hidewidth -\kern-.3em-\hidewidth\cr$z$\cr}}%
}

\begin{document}


\title{Bimodal phase transition in a periodically modulated $\Lambda$-type three-level system}
\author{Sanjoy Mishra}%
\affiliation{Department of Physics and Astronomy, National Institute of Technology Rourkela, Rourkela—769008, India}%

\author{Shraddha Sharma}
 \email{sharmas@nitrkl.ac.in, shrdha1987@gmail.com}
 \affiliation{Department of Physics and Astronomy, National Institute of Technology Rourkela, Rourkela—769008, India}

 \author{Amit Rai}
\affiliation{School of Physical Sciences, Jawaharlal Nehru University, New Delhi 110067, India }

\author{Pitamber Mahanandia}
 \email{pitam@nitrkl.ac.in}
 \affiliation{Department of Physics and Astronomy, National Institute of Technology Rourkela, Rourkela—769008, India}





\date{\today}

\begin{abstract}
We present a theoretical investigation of dynamical quantum phase transitions (QPTs) in a periodically driven $\Lambda$-type three-level system (3LS) embedded in a double-mode cavity, described by a three-level Jaynes-Cumming (3L-JC) Hamiltonian. To begin with, we probe the undriven static Hamiltonian in the dressed-state basis to identify and define distinct coupling regimes and critical points associated with both cavity modes. Furthermore, to investigate the dynamical QPTs in this system, we incorporate a periodic modulation across two atomic states (denoted by $|3\rangle_{at}$ and $|2\rangle_{at}$) out of the three available energy levels. By performing necessary transformations and approximations, we reduce the overall Hamiltonian, which contains static and dynamic modulation terms, into an effective 3L-JC Hamiltonian whose system parameters are dependent on the driving parameters. The validity of our approximations is verified using the Loschmidt echo of time-evolved states corresponding to Hamiltonians before and after the approximations. Finally, we demonstrate that by tuning the modulation parameters, it is possible to explore bimodal superradiant phases in a three-level $\Lambda$-type system while remaining within the critical coupling limits of the static Hamiltonian. Our results provide an insight into the manipulation of quantum phases in a three-level system within an effective extended Jaynes-Cummings regime.

\end{abstract}


\maketitle

\section{Introduction}

Quantum phase transitions (QPTs) \cite{Sachdev2011} in cavity-QED \cite{Landig2016,Hruby2018,Sharma2022,Liu2018,Himbert2019} and circuit-QED \cite{Zhang2014,Nataf2010} systems have gained significant attention lately, owing to a series of discoveries of exotic phases of matter in quantum optical systems, such as superfluid phases \cite{Fisher1989,Greiner2002,Frerot2016}, Mott insulator \cite{Greiner2002,Dickerscheid2003,Kamide2013, KangJun2015}, and ferroelectric phases \cite{Ashida2020} that were previously confined only to condensed matter systems. In addition to these, collective quantum optical effects such as superradiant phase transitions \cite{Hepp1973, Dasgupta2015} and photon blockade phase transition \cite{Fink2017, Curtis2021} have found immediate applications in quantum sensing \cite{Chu2021, Zhu2023, DiCandia2023}, quantum metrology \cite{Garbe2020, DiCandia2023}, and single-photon generation \cite{Zou2020, garbeChen2022}, respectively.
Most of the above described phenomena in the quantum optical systems consist of either large number of emitters in a single cavity or single emitters each embedded to multiply interacting cavities, i.e., the critical phenomena become possible only due to the collective, cooperative behavior of a macroscopically large number of degrees of freedom $N\rightarrow \infty$ (referred to as thermodynamic limit). A classic example of this is the Dicke model \cite{Dicke1954, Wang1973, Hioe1973, Hepp1973, Baden2014}, which is a well-known quantum optical model long known for exhibiting QPT in the thermodynamic limit. {However, in some of the recent studies, it has been demonstrated that QPT and critical phenomena can also be observed in systems composed of only a few interacting components/subsystems, prominently referred to as finite-component/finite-size systems.} In such systems, criticality can appear within a suitable parameter range while keeping a limited number of components \cite{Hwang15, Garbe2020, Yamamoto2025}. They offer greater advantages compared to systems in the thermodynamic limit since the likelihood of fully controlled quantum dynamics in a finite-sized quantum critical system is enhanced owing to a higher degree of coherence \cite{Puebla2017}. Latest works on such systems include QPT in quantum Rabi model (QRM)\cite{Ashhab2010,Ashhab2013,Hwang15,Chen2024}\cite{Puebla2017}, and Jaynes-Cummings systems (JC)\cite{Hwang2016, ChengLiu2023}, to name a few. The thermodynamic limit in such systems is determined by the ratio of atomic transition frequency $\omega$ and cavity mode $\Omega$, i.e $\omega/\Omega\rightarrow \infty$(valid for both two level (2LS) \cite{Bakemeier2012,Ashhab2013} as well as the three level systems (3LS) \cite{Zhang2020, Zhang2025}), thus putting unphysical limits on the experimental conditions.
Consequently, in order to find an alternative to surpass such stringent conditions, dynamic QPT using an external modulating drive have been explored. However, most of these works have been limited to 2LS, where it is fairly direct to obtain simple analytic solutions under minimal approximations \cite{Li2014, Huang2017, Xie2025}. Nevertheless, in most practical scenarios, optical systems possess three or more higher energy levels rather than just 2LS, where approximation conditions could easily break \cite{Veyron2022, Cheng2003}. Previous studies on phase transitions in finite size system, specifically 3LS are in a nascent stage \cite{Robles2020, Zhang2020, Castanos2022, Fan2023}, although lately there is a growing trend in this topic due to its emerging applications in qutrits\cite{Tan2018, Kazmina2024}, quantum sensing \cite{Wang2024}, and quantum batteries\cite{Yang2024}. Investigating phase transitions in such systems offers intrinsic advantages such as tunable transition lifetimes, efficient preparation of atomic states, and longer preservation of coherence, to name a few \cite{Robert2024, Roa2007}. As a result, it becomes necessary to study QPT in multi-level systems where 3LS would serve as a good starting point.

 In this paper, we explore critical coupling limits of a three-level (3L) $\Lambda$-type system coupled to a double-mode cavity described by an extended JC model (i.e 3L-JC) that represents a finite-component system. Our interest lies in probing the QPT that arises when a part of this 3L-JC system is subjected to an external periodic modulation. In order to achieve this objective, we perform appropriate transformations on the periodically modulated Hamiltonian. Subsequent approximations yield an effective 3L-JC Hamiltonian consisting of modified frequencies and coupling strength terms that are now functions of the driving parameters (i.e., driving frequency and amplitude). This indicates that the application of modulating drive alters the system parameters, such as atomic transition frequencies, cavity frequencies and coupling strengths. In this work, we therefore demonstrate that by adjusting these external driving parameters, the ratio of the effective coupling strengths to the effective cavity and atomic frequencies can be tuned beyond the critical coupling limits, achieving 100 times the original coupling strengths. This allows for the observation of exotic bimodal dressed state phases even when the static system parameters are two orders of magnitude less than the critical coupling values. 


\section{Model Hamiltonian \& Equilibrium Phases}

\begin{figure}[htbp]
\centering
\includegraphics[scale=0.5]{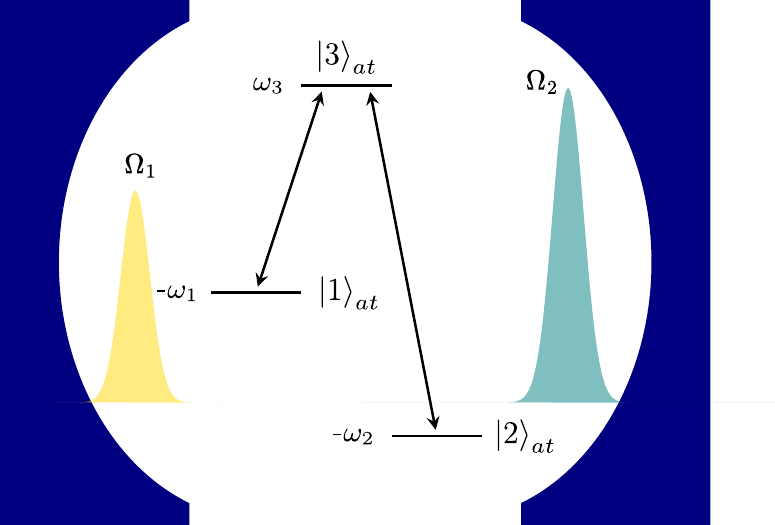}
\caption{Schematic diagram showing a three-level $\Lambda$-type system in a double mode cavity sustaining two resonant frequencies $\Omega_{1}$ and $\Omega_{2}$ respectively. The cavity is assumed to be an ideal lossless cavity $\kappa=0$.}
\label{fig:1}
\end{figure}

The model under consideration involves a 3L atom placed inside a cavity setup, as depicted in Fig.~\ref{fig:1}, where the atomic states are represented as $\ket{1}_{at}$, $\ket{2}_{at}$ and $\ket{3}_{at}$ while $\Omega_{1}$, $\Omega_{2}$ denotes the frequencies of the two resonant cavity modes. Each of these photonic modes couples to a specific atomic transition i.e $\Omega_{1}$ to $\ket{3}_{at}\leftrightarrow\ket{1}_{at}$ and  $\Omega_{2}$ to $\ket{3}_{at}\leftrightarrow\ket{2}_{at}$. Furthermore, these atomic states  $\ket{1}_{at}$, $\ket{2}_{at}$ and $\ket{3}_{at}$ are shown with their corresponding energies $-\omega_{1}$, $-\omega_{2}$ and $\omega_{3}$ respectively. The negative signs are introduced to rescale the mean of the atomic energies to zero \cite{SEN2012224, Nath2008}. Thus the described model in Fig.~\ref{fig:1} is represented by the following Hamiltonian:

\begin{eqnarray}\label{eq:1}
 H &=& \omega_{1}(\sigma_{33}-\sigma_{11}) + \omega_{2}(\sigma_{33}-\sigma_{22}) 
    + \Omega_{1}a_{1}^{\dagger}a_{1}+ \Omega_{2}a_{2}^{\dagger}a_{2} \nonumber\\
    &~&+ g_{1}(\sigma_{31}+\sigma_{13})(a_{1} + a_{1}^{\dagger}) 
    + g_{2}(\sigma_{32}+\sigma_{23})(a_{2} + a_{2}^{\dagger}).\nonumber\\
\end{eqnarray}
 Here, $\sigma_{kj}=|k\rangle\langle j|$, $k,j\in1,2,3$ are atomic transition (for $k\neq j$) or projection operators (for $k=j$). Whereas, $a_{i}$ ($a_{i}^\dagger$) for $i\in(1,2)$ are the annihilation (creation) operators associated with two cavity modes (1, 2). Further, the atom-cavity interaction is denoted by parameters $g_1$ representing $1\leftrightarrow3$ transition coupled to cavity mode-1, while $g_2$ corresponds to the coupling of cavity mode-2 with $2\leftrightarrow3$ transition. The Hamiltonian above is the QRM for a 3LS \cite{SEN2012224, Nath2008}, describing coherent interaction between the electromagnetic field and an atom. This is reflected by the inclusion of counter-rotating part in the last two terms of the Hamiltonian, representing the atom-cavity interaction. 

\begin{figure}[htbp] 
    \centering    
    \includegraphics[width=0.5\textwidth]{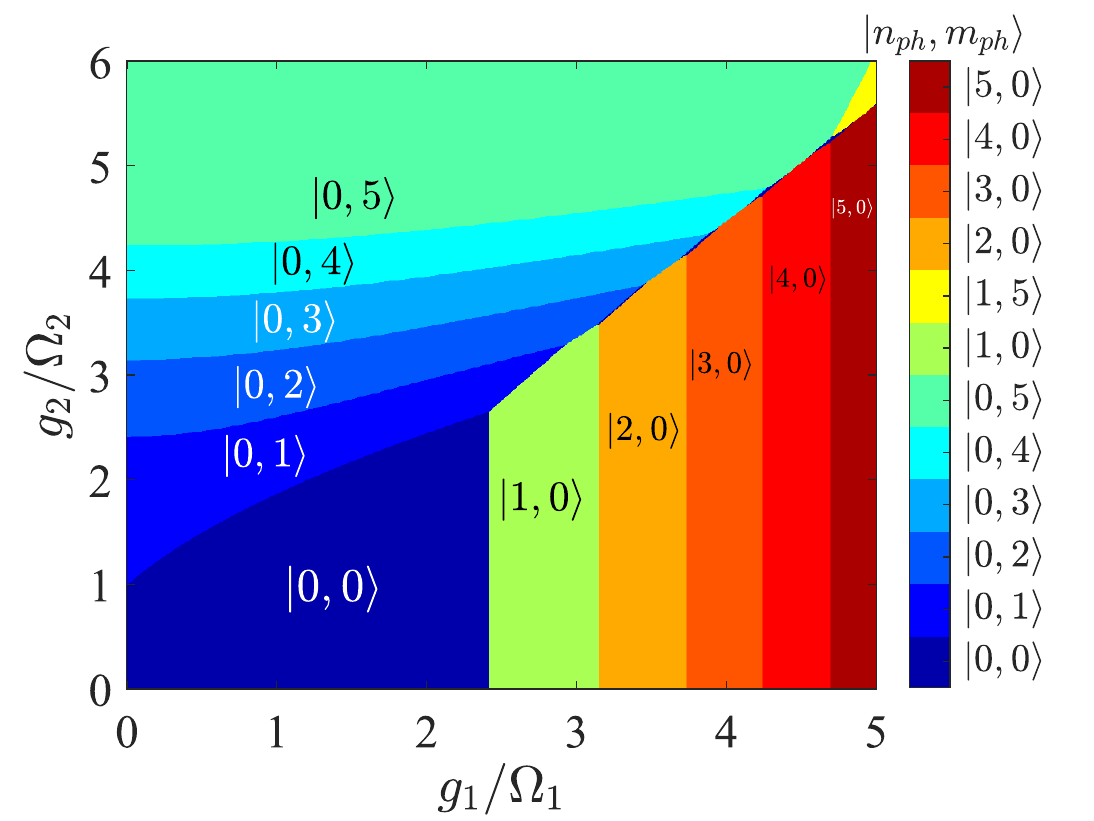} 
    \caption{Numerically computed ground state phase diagram for Hamiltonian in Eq.~\ref{eq:2}, as a function of $g_{1}/\Omega_{1}$ and $g_{2}/\Omega_{2}$ where we have taken $\omega_{1}=0.5$, $\omega_{2}=0.25$, $\delta_1=\delta_2=0$, $\Omega_{1} = 1.25$, and $\Omega_{2} = 1$. Here, $\ket{n_{ph},m_{ph}}$ shows bimodal photon number states depicting different phases.}
    \label{fig:2}
\end{figure}

In this work, we aim to achieve phase transition in the 3L-JC model; therefore, we implement the rotating wave approximation to Eq.~\ref{eq:1} where counter-rotating terms are neglected, thus reducing the 3L-QRM to 3L-JC Hamiltonian, given by: 
\begin{eqnarray}\label{eq:2}
 H_{JC} &=& \omega_{1}(\sigma_{33}-\sigma_{11}) + \omega_{2}(\sigma_{33}-\sigma_{22}) 
    + \Omega_{1}a_{1}^{\dagger}a_{1}+ \Omega_{2}a_{2}^{\dagger}a_{2}\nonumber \\
    &~&+ g_{1}(\sigma_{31}a_1+\sigma_{13}a_1^\dagger) 
    + g_{2}(\sigma_{32}a_2+\sigma_{23}a_2^\dagger).
\end{eqnarray}

\noindent In order to explore different phases arising in this model, we need to perform the diagonalization of the above Hamiltonian. Reasonably, the basis of the above Hamiltonian takes the tensor form $|k\rangle_{at}\otimes|n_{ph}\rangle\otimes|m_{ph}\rangle$ where $|k\rangle$  represents atomic levels $k\in(1,2,3)$ and $|n_{ph}\rangle$ and $|m_{ph}\rangle$ belong to the Fock states for the two modes of the cavity associated with frequencies $\Omega_1$ and $\Omega_2$ respectively. As shown in detail in [\onlinecite{delRio-Lima_2024}], the Hamiltonian in Eq.~\ref{eq:2} can connect to only three distinct basis states; therefore, the whole Hamiltonian in dressed-state basis becomes a $3\times3$ dimensional matrix. In our case, we choose these states to be:
\begin{align} \label{eq:3}
\ket{\psi^{1}} & = \ket{3}_{at} \ket{n_{ph}} \ket{m_{ph}-1},\nonumber\\
    \ket{\psi^{2}} & = \ket{1}_{at} \ket{n_{ph}+1} \ket{m_{ph}-1}, \nonumber \\
    \ket{\psi^{3}} & = \ket{2}_{at} \ket{n_{ph}} \ket{m_{ph}} 
\end{align}

\noindent where the subscript `$at$' as already discussed previously, signifies the atomic state, differentiating it from the two other photonic states represented in the Fock/number state basis. A comprehensive expression for the Hamiltonian is provided in Appendix \ref{Appndx:Appendix A}. Furthermore, Eq.~\ref{hnm} represents the Hamiltonian in the dressed-state basis, which, upon diagonalization yields the ground state energy phase diagram.  Subsequently, we construct a pseudocolor plot of ground state energy as a function of the two coupling parameters, $g_1/\Omega_1$ and $g_2/\Omega_2$, characterizing the system as shown in Fig.~\ref{fig:2}. Distinct colormaps reveal that the critical transition from the $\ket{0,0}$ photon state to higher-excitation in mode-1 ($n_{ph}$) and mode-2 ($m_{ph}$) of the cavity occurs at well-defined coupling ratios $g_1/\Omega_1$ and $g_2/\Omega_2$. The ground state acquires non-zero photon occupation only across these sharply defined boundaries in parameter space indicating an equivalent superradiant phase transition for 3LS. To further simplify our understanding of bimodal superradiant states arising in the phase diagram, we classify them using the following nomenclatures, normal-phase for $\ket{0,0}$ state, $y_{1}$-type for $\ket{n_{ph},0}$, $y_{2}$-type for $\ket{0,m_{ph}}$ and lastly mixed-phase for $\ket{n_{ph},m_{ph}}$ \cite{Zhang2020}. 
Keeping these terminologies in mind, if we vary $g_{2}/\Omega_{2}$ beyond critical coupling value while simultaneously keeping $g_{1}/\Omega_{1}$ fixed within $0<g_{1}/\Omega_{1}<2.41$ range, we observe only $y_{2}$-type ground state phase transitions $\ket{0,0}\rightarrow\ket{0,1}\rightarrow\ket{0,2}\rightarrow\ket{0,3}\rightarrow\ket{0,4}\rightarrow\ket{0,5}$. Similarly by fixing $g_{2}/\Omega_{2}$ within $0<g_{2}/\Omega_{2}<1$ and varying $g_{1}/\Omega_{1}$ beyond $2.41$, we get only $y_{1}$-type ground state phase transition $\ket{0,0}\rightarrow\ket{1,0}\rightarrow\ket{2,0}\rightarrow\ket{3,0}\rightarrow\ket{4,0}\rightarrow\ket{5,0}$. A combination of  $y_{1}$-type and $y_{2}$-type phase transition ($\ket{0,1}\rightarrow\ket{0,0}\rightarrow\ket{1,0}\rightarrow\ket{2,0}\rightarrow\ket{3,0}\rightarrow\ket{4,0}\rightarrow\ket{5,0}$, or $\ket{0,2}\rightarrow\ket{0,1}\rightarrow\ket{0,0}\rightarrow\ket{1,0}\rightarrow\ket{2,0}\rightarrow\ket{3,0}\rightarrow\ket{4,0}\rightarrow\ket{5,0}$) also arises for $1<g_{2}/\Omega_{2}<2.64$ while changing $g_{1}/\Omega_{1}$ from 0 to beyond $2.41$. Furthermore, if we move beyond $(g_{1}/\Omega_{1}, g_{2}/\Omega_{2}) =(2.41, 2.64)$, we start observing ground states of mixed-phase type in combination with $y_{1}$-type and $y_{2}$-type phases. It is evident from Fig.~\ref{fig:2} that $\ket{1,5}$ belongs to the category of mixed phase. However, it should be noted that many such mixed phases appear along the diagonal (separating $y_{1}$-type and $y_{2}$-type phases) in extremely small regions beyond $(g_{1}/\Omega_{1}, g_{2}/\Omega_{2}) =(2.41, 2.64)$. Although these are not clearly visible in Fig.~\ref{fig:2}, owing to the rescaling of the colorbar carried out for simplicity.


Thus, to summarize, in order to observe a transition from the normal phase $|0,0\rangle$ to any of the superradiant phases ($y_{1}$, $y_{2}$ and mixed-type), the critical coupling strengths must satisfy the following conditions: if $g_1/\Omega_1$ is fixed at some value less than 2.415, then $g_{2}/\Omega_{2}$ must be greater than 1; conversely, if $g_2/\Omega_2$ is fixed below 1, then $g_{1}/\Omega_{1}$ must be no less than 2.415.  However, exceeding these values for $g_{1}$ or $ g_{2}$ would correspond to taking the system into a non-perturbative regime where the rotating wave approximation (RWA) and hence (Eq.~\ref{eq:2}) would no longer be valid \cite{DeBernardis2024}. {Thus, the primary objective of our study is to investigate a regime wherein an equivalent superradiant phase emerges at reduced coupling strengths, thereby facilitating the observation of QPTs in a 3L-JC Hamiltonian within a lower parameter range. To achieve this, we introduce a sinusoidal modulation to the Hamiltonian in Eq.~\ref{eq:2} and derive an effective Hamiltonian. This framework would allow us to systematically analyze phase transitions in an extended JC model tailored for $\Lambda$-type 3LS.}

\section{Periodic modulation and effective Hamiltonian}
As mentioned in the previous section, we include a driving term along with the Hamiltonian in Eq.~\ref{eq:2}, given by,
\begin{equation}\label{eq:4}
H_{D}(t) = A_{D}cos(\omega_{D}t)(\sigma_{33}-\sigma_{22}),
\end{equation}
\noindent where, $A_D$ is the amplitude of the modulation, $\omega_D$ is the driving frequency, and, $t$ is the time. We then perform a unitary transformation of the combined Hamiltonian, $H(t)=H_{JC}+H_D(t)$, using unitary operator $U_{1}$ given by,
\begin{eqnarray}\label{eqn:5}
U_{1}(t) &=& \mathcal{T} \exp\Bigl(-i\int\Bigl\{-\omega_{1}\sigma_{11}- \omega_{2}\sigma_{22} +\Omega_{1}a_{1}^{\dagger}a_{1}+ \Omega_{2}a_{2}^{\dagger}a_{2}\nonumber  \\
&~&+ \frac{A_{D}}{\omega_{D}}sin(\omega_{D}t)(\sigma_{33}-\sigma_{22})\Bigl\}dt\Bigl),  
\end{eqnarray}

\noindent{where, $\mathcal{T}$ denotes the time-ordering operator. This transformation $U_1^\dagger(t)H(t)U_1(t)+i\dot{U}_1^\dagger(t)U_1(t)$ changes the Hamiltonian from the lab frame to the rotated frame of the external driving. After applying this transformation, the new Hamiltonian is given by,}

\begin{eqnarray}\label{eq:6}
H_{rot}(t) &=& g_{1}J_{0}(\theta)\Bigl[ \sigma_{31}a_{1}e^{i\delta_1 t} + \sigma_{13}a_{1}^{\dagger}e^{i\delta_1 t} \Bigr]\nonumber\\
&+& g_{1}\Bigl[ \sigma_{31}a_{1}\sum_{\substack{p = -\infty \\ p \neq 0}}^{\infty} J_{p}(\theta)e^{ip\omega_{D}t}e^{i\delta_1 t}\nonumber\\
&+& \sigma_{13}a_{1}^{\dagger}\sum_{\substack{p = -\infty \\ p \neq 0}}^{\infty} J_{p}(\theta)e^{-ip\omega_{D}t}e^{-i\delta_1 t}\Bigr] \nonumber \\ 
&+& g_1J_{n_0}(\theta)\Bigl[\sigma_{31}a_{1}^{\dagger}e^{i\Delta_{n_0}t} + \sigma_{13}a_{1}e^{-i\Delta_{n_0}t}\Bigr] \nonumber \\ 
&+& g_1\Bigl[ \sigma_{31}a_{1}^{\dagger}\sum_{\substack{n = -\infty \\ n \neq n_{0}}}^{\infty}J_{n}(\theta)e^{i\Delta_{n}t}\nonumber\\
&+& \sigma_{13}a_{1}\sum_{\substack{n = -\infty \\ n \neq n_0}}^{\infty}J_{n}(\theta)e^{-i\Delta_{n}t}\Bigr] \nonumber \\ 
&~& + g_{2}J_{0}(2\theta)\Bigl[ \sigma_{32}a_{2}e^{i\delta_2 t} + \sigma_{23}a_{2}^{\dagger}e^{-i\delta_2 t}\Bigr]  \nonumber \\ 
&~& + g_{2}\Bigl[ \sigma_{32}a_{2}\sum_{\substack{q = -\infty \\ q \neq 0}}^{\infty} J_{q}(2\theta)e^{iq\omega_{D}t}e^{i\delta_2 t} \nonumber \\ 
&~& \quad + \sigma_{23}a_{2}^{\dagger}\sum_{\substack{q = -\infty \\ q \neq 0}}^{\infty} J_{q}(2\theta)e^{-iq\omega_{D}t}e^{-i\delta_2 t}\Bigr] \nonumber \\ 
&~& + g_2J_{m_0}(2\theta)\Bigl[\sigma_{32}a_{2}^{\dagger}e^{i\Delta_{m_0}t} + \sigma_{23}a_{2}e^{-i\Delta_{m_0}t}\Bigr] \nonumber \\ 
&+& g_2\Bigl[ \sigma_{32}a_{2}^{\dagger}\sum_{\substack{m = -\infty \\ m \neq m_{0}}}^{\infty}J_{m}(2\theta)e^{i\Delta_{m}
t}\nonumber\\
&+& \sigma_{23}a_{2}\sum_{\substack{m = -\infty \\ m \neq m_0}}^{\infty}J_{m}(2\theta)e^{-i\Delta_{m}t}\Bigr].
\end{eqnarray}

\noindent{Here, $\theta=A_D/\omega_D$, represents the ratio of driving amplitude to driving frequency. Whereas, $\delta_1$, $\delta_2$, $\Delta_1$ and $\Delta_2$, in terms of the system parameters, are defined as,}
\begin{align}\label{eq:7}
    \delta_{1} &= 2\omega_1 + \omega_2 - \Omega_1, \nonumber \\
    \delta_{2} &= 2\omega_2 + \omega_1 - \Omega_2 
\end{align}
\noindent and,
\begin{align}
    \Delta_{n} &= 2\omega_1 + \omega_2 + \Omega_1 + n\omega_D, \nonumber \\
    \Delta_{m} &= 2\omega_2 + \omega_1 + \Omega_2 + m\omega_D. \label{eq:8}
\end{align}
\noindent The $\delta_{1}$ and $\delta_{2}$ in Eq.~\ref{eq:7}  represents the cavity detunings reflecting the frequency mismatch between atomic transitions ($\ket{3}_{at}\leftrightarrow\ket{1}_{at}$, $\ket{3}_{at}\leftrightarrow\ket{2}_{at}$) and corresponding coupled cavity modes ($1$, ${2}$, respectively). Whereas, the $J_{l\in(p,n,q,m)}$ are  Bessel function of $l$th order, introduced via Jacobi-Anger identity \cite{Xie2025},
\begin{equation}
     {\rm e}^{i\zstroke\sin{\theta}} = \sum_{l=-\infty}^{\infty} J_{l}(\zstroke)e^{il\theta},\label{eq:9}
\end{equation}

\noindent which we have applied to obtain $H_{rot}(t)$ in Eq.~\ref{eq:6}. These $p, n, q, m\in\mathbb{Z}$, $\mathbb{Z}$ being the set of all integers. It can be observed from the Hamiltonian $H_{rot}(t)$ that the new coupling strengths in the rotating frame ($g_{1}J_n(\theta)$ and $g_{2}J_m(2\theta)$), and the oscillating frequency of the side band in rotating terms ($\delta_1+p\omega_D$, $\delta_2+q\omega_D$) as well as in the counter rotating terms ($\Delta_n$ and $\Delta_m$), all depends on the modulation amplitude, $A_D$ and the frequency $\omega_D$, which can be tuned accordingly. In Eq.~\ref{eq:6}, $n_0, m_0$ are introduced, in order to minimize $\Delta_n$ and $\Delta_m$, respectively. This is performed to ensure that we can distinctly separate out fast-rotating terms ($\Delta_{n\neq n_0}$, $\Delta_{m\neq m_0}$) from the minimized $\Delta_{n_0}$, $\Delta_{m_0}$. Therefore, $\Delta_{n_0}$ and $\Delta_{m_0}$ are given by expressions, 
\begin{align}\label{eq:10}
    \abs{\Delta_{n_0}} &= \min{\Delta_{n}} = \abs{2\omega_1 + \omega_2 + \Omega_1 + n_0\omega_D}, \nonumber \\
    \abs{\Delta_{m_0}} &= \min{\Delta_{m}} = \abs{2\omega_2 + \omega_1 + \Omega_2 + m_0\omega_D}.
\end{align}




\noindent Further, we can impose the following conditions on $\delta_{1}$, $\delta_{2}$, $g_{1}$ and $g_{2}$ in order to simplify $H_{rot}$,
\begin{align} \label{eq:11}
    \omega_{D} &\gg \abs{\delta_{1}}, \abs{\delta_{2}}, \nonumber \\
    \omega_{D} &\gg \abs{\Delta_{n_{0}}}, \abs{\Delta_{m_{0}}},  \nonumber \\
    \omega_{D} &\gg g_{1} > g_{1} \abs{J_{n}(\theta)},  \nonumber \\
    \omega_{D} &\gg g_{2} > g_{2} \abs{J_{m}(2\theta)}.
\end{align}
\noindent The conditions above removes the fast oscillating terms in $H_{\textit{rot}}$ and reduces it to a Hamiltonian referred to as $H_{\textit{eff}}$, given by:-
\begin{align} \label{eq:12}
    H_{\textit{eff}}\approx & g_{1}J_{0}(\theta)\Bigl[ \sigma_{31}a_{1}e^{i\delta_1 t} + \sigma_{13}a_{1}^{\dagger}e^{i\delta_1 t} \Bigr] \nonumber \\
    & + g_1J_{n_0}(\theta)\Bigl[\sigma_{31}a_{1}^{\dagger}e^{i\Delta_{n_0}t} + \sigma_{13}a_{1}e^{-i\Delta_{n_0}t}\Bigr] \nonumber \\ 
    & + g_{2}J_{0}(2\theta)\Bigl[ \sigma_{32}a_{2}e^{i\delta_2 t} + \sigma_{23}a_{2}^{\dagger}e^{-i\delta_2 t}\Bigr]  \nonumber \\ 
    & + g_2J_{m_0}(2\theta)\Bigl[\sigma_{32}a_{2}^{\dagger}e^{i\Delta_{m_0}t} + \sigma_{23}a_{2}e^{-i\Delta_{m_0}t}\Bigr]. 
\end{align}

We now apply the second unitary transformation $U_2$, which in terms of the effective cavity mode frequencies $\tilde{\Omega}_{1}$ and $\tilde{\Omega}_{2}$ and effective frequencies of the atomic states $\ket{1}_{at}$ and $\ket{2}_{at}$ given by $\tilde{\omega}_{1}$ and $\tilde{\omega}_{2}$, respectively is,
\begin{align} \label{eq:u2}
    U_2 = e^{i\tilde{\Omega}_1a_{1}^{\dagger}a_{1}t}e^{i\tilde{\Omega}_2a_{2}^{\dagger}a_{2}t}e^{i\tilde{\omega}_1\sigma_{33}t}e^{-i\tilde{\omega}_1\sigma_{11}t}e^{i\tilde{\omega}_2\sigma_{33}t}e^{-i\tilde{\omega}_2\sigma_{22}t},
\end{align}

\noindent where,

\begin{eqnarray}
  \tilde{\Omega}_{1} &=&\dfrac{\Delta_{n_0} - \delta_1}{2}, \; \tilde{\Omega}_{2} = \dfrac{\Delta_{m_0} - \delta_2}{2}\label{tild_Omeg}\\
     \tilde{\omega}_{1} &=&\dfrac{(2\delta_1 - \delta_2)+(2\Delta_{n_0}-\Delta_{m_0})}{6},\nonumber \\
    \tilde{\omega}_{2} &=&\dfrac{(2\delta_2 - \delta_1)+(2\Delta_{m_0}-\Delta_{n_0})}{6}.\label{tild_omeg}
\end{eqnarray}


\noindent Therefore, the new effective Hamiltonian $\tilde{H}_{\textit{eff}}$ after applying $U_2$ on $H_{\textit{eff}}$ comes out to be,
\begin{eqnarray}\label{eq:13}
    \tilde{H}_{\textit{eff}}&=& \tilde{\omega}_{2}(\sigma_{33}-\sigma_{22})+\tilde{\omega}_{1}(\sigma_{33}-\sigma_{11})+\tilde{\Omega}_{2}a_{2}^{\dagger}a_{2}+\tilde{\Omega}_{1}a_{1}^{\dagger}a_{1}\nonumber \\       &+&g_{r_{1}}\Big[\sigma_{31}a_{1}e^{i\delta_1 t}e^{i\tilde{\Omega}_{1} t}e^{-i2\tilde{\omega}_{1} t}e^{-i\tilde{\omega}_{2} t} + h.c\Big]\nonumber \\    &+&g_{c_{1}}\Big[\sigma_{31}a_{1}^{\dagger}e^{i\Delta_{n_0}t}e^{-i\tilde{\Omega}_{1}t}e^{-i2\tilde{\omega}_{1} t}e^{-i\tilde{\omega}_{2} t} + h.c\Big]\nonumber \\    &+&g_{r_{2}}\Big[\sigma_{32}a_{2}e^{i\delta_2 t}e^{i\tilde{\Omega}_{2} t}e^{-i\tilde{\omega}_{1} t}e^{-i2\tilde{\omega}_{2} t} + h.c\Big]\nonumber \\    &+&g_{c_{2}}\Big[\sigma_{32}a_{2}^{\dagger}e^{i\Delta_{m_0}t}e^{-i\tilde{\Omega}_{2}t}e^{-i\tilde{\omega}_{1} t}e^{-i2\tilde{\omega}_{2} t} + h.c\Big],
\end{eqnarray}
\noindent in the above equation, $g_{r_{1}}$, $g_{r_{2}}$, $g_{c_{1}}$, $g_{c_{2}}$ are defined as,

\begin{eqnarray}
     g_{r_{1}}(\theta) &=& g_{1}J_{0}(\theta), ~~~~
    g_{r_{2}}(2\theta) = g_{2}J_{0}(2\theta) \label{gr12} \\
    g_{c_{1}}(\theta) &=& g_{1}J_{n_{0}}(\theta), ~~~~
    g_{c_{2}}(2\theta) = g_{2}J_{m_{0}}(2\theta).\label{gc12}
\end{eqnarray}

 \noindent{The terms $g_{r_{1}}$ and $g_{r_{2}}$ represents the effective coupling strengths at the zeroth order sideband while $g_{c_{1}}$ and $g_{c_{2}}$ represents the effective coupling strengths at the $n_{0}$-th and $m_{0}$-th order sideband respectively \cite{Li2025}. It is important to note that the expressions for $\tilde{\Omega}_{1}$, $\tilde{\Omega}_{2}$, $\tilde{\omega}_{1}$, and $\tilde{\omega}_{2}$ in Eq. \ref{tild_Omeg}, and \ref{tild_omeg}, respectively, are chosen such that the exponents in Eq.~\ref{eq:13} vanish. This condition ensures the elimination of the explicit time dependence in the effective Hamiltonian $\tilde{H}_{\textit{eff}}$, leading to $\tilde{H}_{\textit{\textit{eff}}_{1}}$,}
 
 \begin{eqnarray}\label{eq:18}
\tilde{H}_{\textit{eff}_{1}}&=& \tilde{\omega}_{2}(\sigma_{33}-\sigma_{22})+\tilde{\omega}_{1}(\sigma_{33}-\sigma_{11})+\tilde{\Omega}_{2}a_{2}^{\dagger}a_{2}+\tilde{\Omega}_{1}a_{1}^{\dagger}a_{1} \nonumber\\       &+&g_{r_{1}}\Big[\sigma_{31}a_{1} + h.c\Big] + g_{c_{1}}\Big[\sigma_{31}a_{1}^{\dagger} + h.c\Big] \nonumber\\    &+&g_{r_{2}}\Big[\sigma_{32}a_{2} + h.c\Big] + g_{c_{2}}\Big[\sigma_{32}a_{2}^{\dagger} + h.c\Big].
\end{eqnarray}

 The effective 3L-JC Hamiltonian can now be retrieved from $\tilde{H}_{\textit{eff}_{1}}$ in Eq.~\ref{eq:18}, if the counter-rotating terms represented by $g_{c_{1}}$ and $g_{c_{2}}$ has negligible contribution as compared to the rest of the terms in the Hamiltonian. This is achieved when the following conditions are satisfied:
\begin{eqnarray} \label{eq:17}
g_{c_{1}}/\Delta_{n0} &\ll&1, ~~~~ 
    g_{c_{2}}/\Delta_{m0} \ll1. 
\end{eqnarray}

 \noindent{Accordingly, we neglect these terms and the final effective 3L-JC Hamiltonian can therefore be represented by, }

 \begin{eqnarray} \label{eq:H3JC}
    \tilde{H}_{3JC}&=& \tilde{\omega}_{2}(\sigma_{33}-\sigma_{22})+\tilde{\omega}_{1}(\sigma_{33}-\sigma_{11})+\tilde{\Omega}_{2}a_{2}^{\dagger}a_{2}\nonumber \\
&+&\tilde{\Omega}_{1}a_{1}^{\dagger}a_{1} +{g_{r_{1}}\Big[\sigma_{31}a_{1} + h.c\Big]}+{g_{r_{2}}\Big[\sigma_{32}a_{2} + h.c\Big]}.\nonumber\\
\end{eqnarray}

\noindent{This final Hamiltonian is now in terms of the new frequencies and interactions between the atom and the cavity owing to the sinusoidal modulation.} It is to be noted that all these effective parameters can be tuned by adjusting the ratio of driving amplitude to frequency, \textit{i.e.} $A_{D}/\omega_{D}=\theta$. Therefore, in the next section, we will target the range of relevant parameters that will serve our purpose of obtaining different out-of-equilibrium phases even in the normal phase region of the undriven (static) Hamiltonian.

\begin{figure}[h]
    \centering
    \begin{subfigure}{0.4\textwidth}
        \centering
        \includegraphics[width=\textwidth]{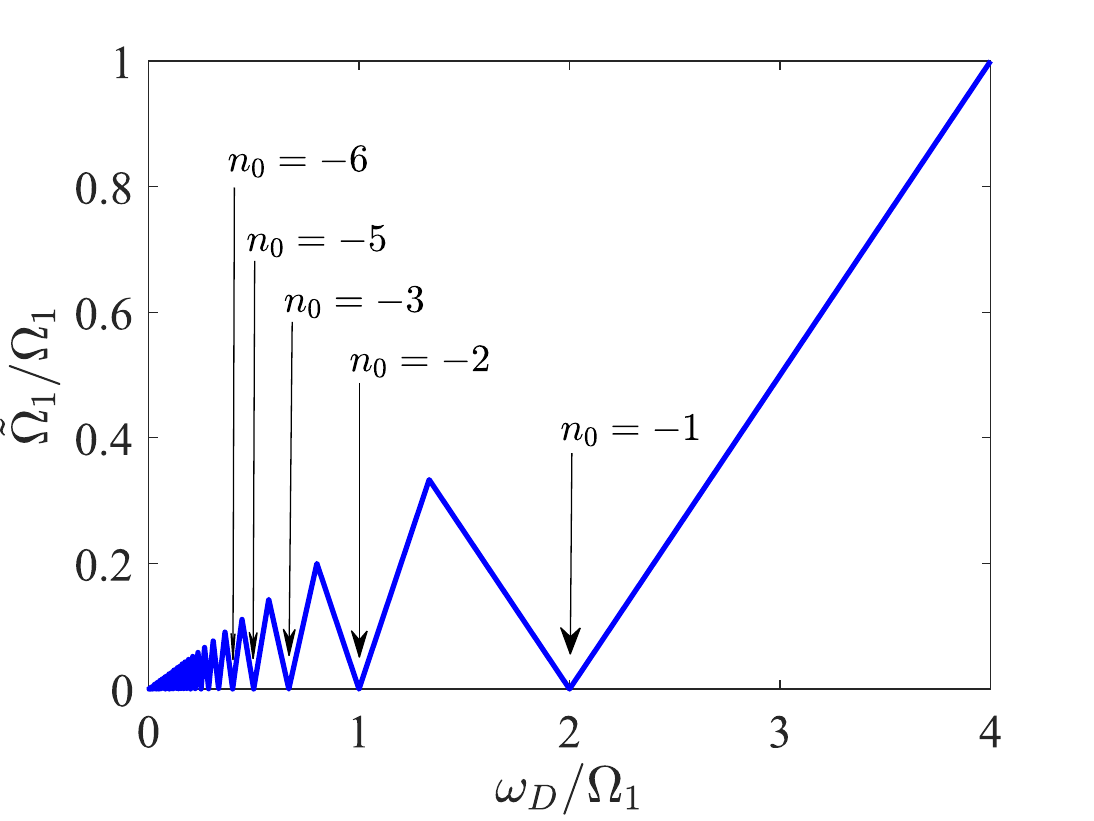} 
        \caption{The plot for $\tilde\Omega1/\Omega_1$ vs $\omega_D/\Omega_1$ showing $V$-shaped valleys, with $n_0$ values indicating minimum value taken by $\tilde\Omega1/\Omega_1$.}
        \label{fig:3a}
    \end{subfigure}

    \vspace{0.06cm} 
    \begin{subfigure}{0.4\textwidth}
        \centering
        \includegraphics[width=\textwidth]{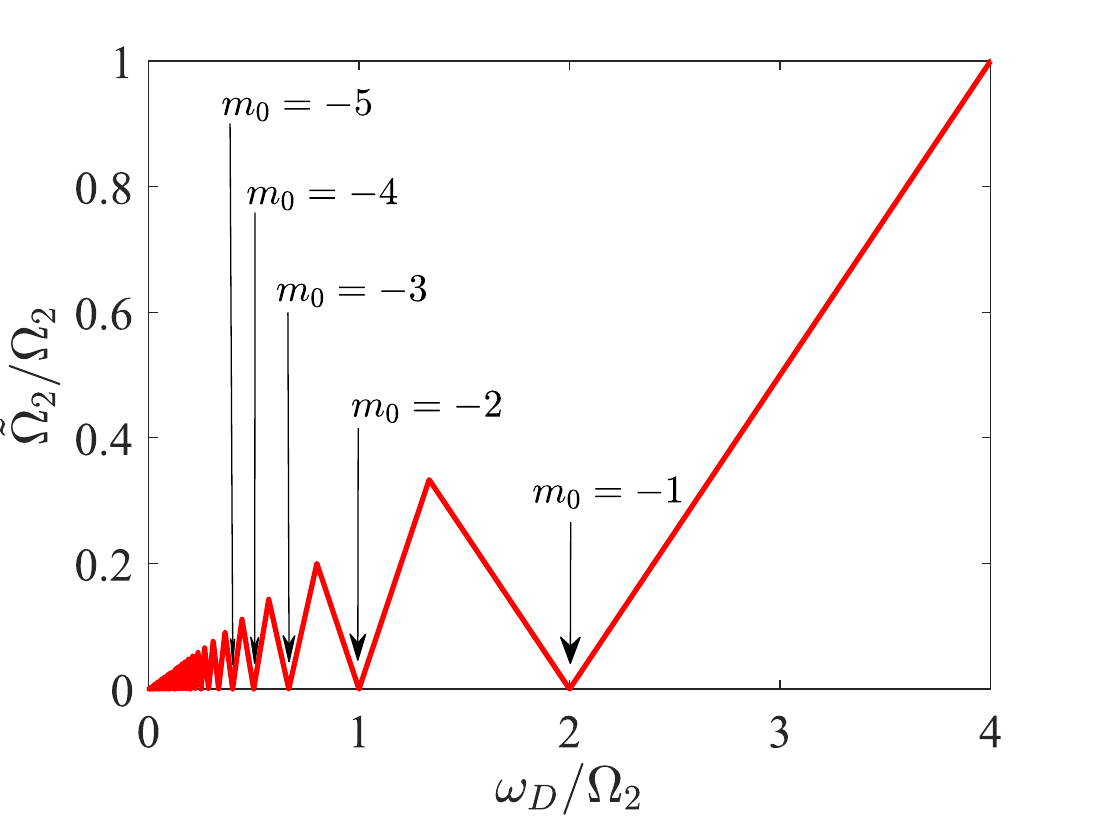}
        \caption{$\tilde\Omega_2/\Omega_2$ vs $\omega_D/\Omega_2$ plot with $m_0$ marking the minima of $V$-shaped valleys.}
        \label{fig:3b}
    \end{subfigure}
\caption{Plots depicting variation of the effective cavity frequency $\tilde{\Omega}_{1(2)}$ w.r.t. the external driving frequency $\omega_D$ with $(a)$ $n_{0}$ and (b) $m_0$ indicating the minima of $\tilde{\Omega}_{1}$ and $\tilde{\Omega}_{2}$ respectively. We have set other parameters to be $\omega_1=0.5$, $\omega_2=0.25$, $\Omega_1=1.25$, $\Omega_2=1$, and $\delta_1=\delta_2=0$.}
    \label{fig:3}
\end{figure}

\section{Relevant Parameter Regime}
In this section, we analyze the behavior of different parameters concerning $\omega_D$ or $\theta$. Starting from Eq.~\ref{tild_Omeg}, we plot $\tilde{\Omega}_{1(2)}/\Omega_{1(2)}$ versus $\omega_D/\Omega_{1(2)}$ in Fig.~\ref{fig:3a} (\ref{fig:3b}) for $\delta_1=\delta_2=0$. One observes V-shaped valleys up until $\omega_D/\Omega_{1(2)}<4$ after which they saturate to 1 (omitted from the figure due to its trivial nature). The zero values of $\tilde{\Omega}_1/\Omega_1$ and $\tilde{\Omega}_2/\Omega_2$ correspond to $n_0$ and $m_0$, respectively, and remain unchanged throughout the entire valley. It can further be shown that these zeros of $\tilde{\Omega}_1/\Omega_1$ and $\tilde{\Omega}_2/\Omega_2$ occur at: 

\begin{eqnarray}
\omega_D&=&\frac{-2(2\omega_1+\omega_2)}{n_0}=\frac{-2\Omega_1}{n_0},~ \text{and},
   \label{omegaD1} \\   \omega_D&=&\frac{-2(2\omega_2+\omega_1)}{m_0}=\frac{-2\Omega_2}{m_0},~ \text{respectively.} \label{omegaD2} 
\end{eqnarray}

\begin{figure}[ht]
    \centering
    \includegraphics[width=0.5\textwidth]{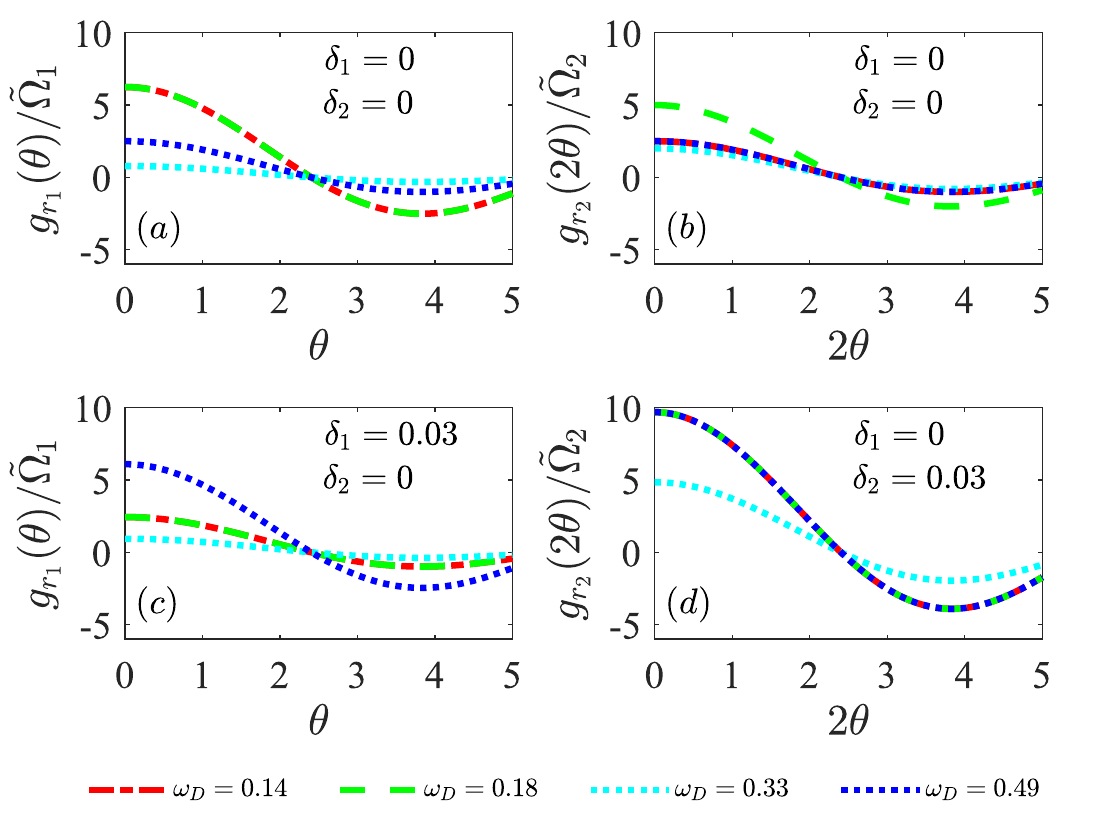} 
    \caption{Plots of $g_{r_{1}}$ and $g_{r_{2}}$ w.r.t $\theta$ and $2\theta$ respectively for different $\omega_D$ and detunings. The sideband $n_0$ and $m_0$ values corresponding to $\omega_{D}=0.14$, $0.18$, $0.33$ and $0.49$ are  $(n_{0},m_{0})=$ $(-17,-14)$, $(-14,-11)$, $(-7,-6)$, $(-5,-4)$, respectively. $\omega_1=0.5$, $\omega_2=0.25$, with coupling strengths set to $g_{1}=g_2=0.05$. For subplots $\boldsymbol{(a)}$ and $\boldsymbol{(b)}$ $\Omega_1=1.25$, $\Omega_2=1$ whereas for $\boldsymbol{(c)}$ and $\boldsymbol{(d)}$ $\Omega_1=1.22$ and $\Omega_2=0.97$ respectively. }
    \label{fig:gr}
\end{figure}

\begin{figure}[ht]
    \centering
    \includegraphics[width=0.5\textwidth]{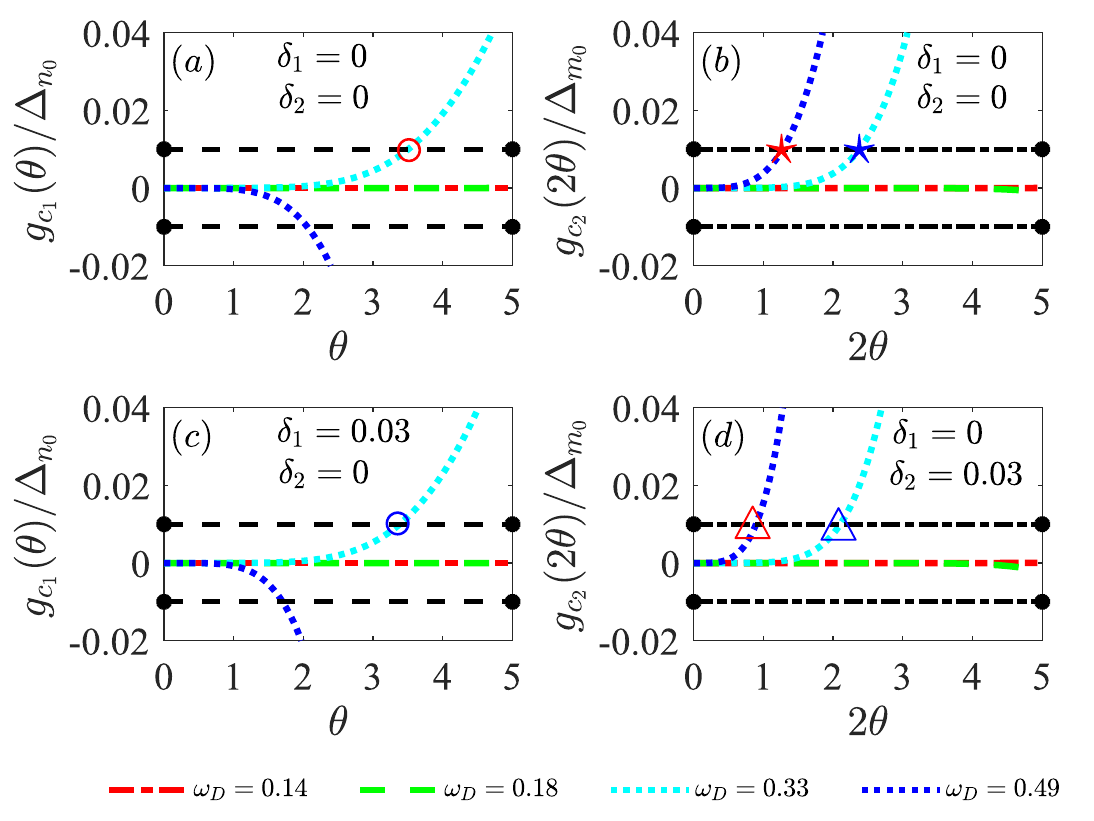} 
    \caption{Variation of $g_{c_{1}}(\theta)$ and $g_{c_{2}}(2\theta)$ w.r.t $\theta$ and $2\theta$ respectively for different detunings and $\omega_D$. The symbols  {\(\circ\)}, {\(\star\)}, and  {\(\triangle\)} are inserted to highlight the positions where the counter-rotating terms become non-negligible. The $\omega_{D}= 0.14$, $0.18$, $0.33$  and  $0.49$ corresponds to sideband values $(n_{0},m_{0})=(-17,-14)$, $(-14,-11)$, $(-7,-6)$, $(-5,-4)$ respectively. The values for $\omega_1$, $\omega_2$, $\Omega_1$, $\Omega_2$ follows the same template as Fig.\ref{fig:gr}. }
    \label{fig:gc}
\end{figure}

Following the investigation on the values of $n_0$ and $m_0$, we can move forward to explore the behavior of effective coupling strengths. Based on Eq.~\ref{gr12}, it is evident that  $g_{r_{1}}$ is a function of $\theta$, while $g_{r_2}$ is a function of $2\theta$. It can be observed in Fig.~\ref{fig:gr} that, as $\theta$ is changed both $g_{r_{1}}$ and $g_{r_2}$ scaled by $\tilde{\Omega}_{1}$ and $\tilde{\Omega}_{2}$, respectively, exhibits an oscillatory behavior. In Fig.~\ref{fig:gr}$(a)$, $(b)$ we have taken $\delta_{1}=\delta_{2}=0$, it was observed, as $\theta$ tends to zero the $g_{r_{1}}/\tilde{\Omega}_{1}$ and $g_{r_{2}}/\tilde{\Omega}_{2}$ increases and surpasses the static Hamiltonian's normal phase region bounded approximately by $g_1/\Omega_1=2.41$ and $g_2/\Omega_2=2.64$. This indicates that the effective coupling strengths can support higher-order superradiant phases as the system is periodically driven. In the case when one of the detuning is kept finite, as shown in Fig.~\ref{fig:gr} $(c)$ and $(d)$, the oscillatory behavior still persists, however the maximum value of $g_{r_1}$ or $g_{r_2}$ for small $\theta$ changes depending on the interplay between the detunings and the driving frequency. It was observed in Fig.~\ref{fig:gr} $(c)$ that the values of $g_{r_{1}}$  at small $\theta$ is enhanced for $\omega_D=0.49$, while it either decreases or remains unchanged for rest of the smaller driving frequencies. Whereas in Fig.~\ref{fig:gr} $(d)$, we show that the maximum value of $g_{r_{2}}$  for the detuning values $\delta_1=0$,~$\delta_2=0.03$, is significantly enhanced irrespective of $\omega_D$ values. On the other hand, the behavior of the interaction strength of the counter-rotating terms, denoted by $g_{c_1}$ and $g_{c_2}$, rescaled w.r.t $\Delta_{n_0}$ and $\Delta_{m_0}$, respectively, is presented in Fig.~\ref{fig:gc}. The expressions for $g_{c_1}$ and $g_{c_2}$ are given by Eq.~\ref{gc12}, where we have used Fig.~\ref{fig:3} to extract the values of $n_0$ and $m_0$. Notably, as  $\omega_D$ is increased, $g_{c_1}/\Delta_{n_0}$ and $g_{c_2}/\Delta_{m_0}$ becomes non-negligible beyond a critical $\theta$ marked by symbols (\textcolor{blue}{\(\circ\)}, \textcolor{red}{\(\circ\)}, \textcolor{blue}{\(\star\)}, \textcolor{red}{\(\star\)}, \textcolor{blue}{\(\triangle\)} and \textcolor{red}{\(\triangle\)}), where these rescaled couplings cross the 0.01 value (denoted by a black dotted line in Fig.~\ref{fig:gc}). It is to be noted that for the Hamiltonian in Eq.~\ref{eq:H3JC} to be valid, $g_{c_1}/\Delta_{n_0}$ and $g_{c_2}/\Delta_{m_0}$, must remain below $0.01$ to fulfill the condition of Eq.~\ref{eq:17}. Therefore, the behavior of counter-rotating terms in Fig.~\ref{fig:gc} provides information on the relevant range of $\theta$ and values of $\omega_D$.


 With these parameter ranges and values taken into account, we will further explore the behavior of ground-state quantum fidelity between different Hamiltonians, verifying the validity of the approximations taken into account to obtain Eq.~\ref{eq:12}, \ref{eq:18}, and \ref{eq:H3JC} in the next section.

\section{Validity of approximations: {Loschmidt Echo}}

{Now that the relevant parameters are obtained, in order to verify our claim that all the above approximations leading to the final effective Hamiltonian holds true, we proceed to evaluate the} {time evolved} {fidelity, also known as Loschmidt echo \cite{Wisniacki2012,Sharma2015,CamposVenuti2010} or the time-evolved fidelity. It can be defined as a quantity which measures the degree of overlap between} {ground state} {  wavefunctions} evolved with two different Hamiltonians, expressed as:-
\begin{equation}
   F(t)=\abs{\braket{\psi_{exact}(t)}{\psi_{approx}(t)}}^{2}.
\end{equation}
\noindent In the above equation, $\ket{\psi_{\text{exact}}}$ denotes the ground state wavefunction of the exact Hamiltonian, either obtained via a unitary transformation (i.e., Eq.~\ref{eq:6} or Eq.~\ref{eq:13}) or corresponding to the Hamiltonians prior to applying any approximations (Eq.~\ref{eq:11} or Eq.~\ref{eq:17}). In contrast, $\ket{\psi_{\text{approx}}}$ represents the ground state wavefunction of the Hamiltonian after the application of the respective approximations (Eq.~\ref{eq:12} or Eq.~\ref{eq:H3JC}) \cite{Sharma2014,Cucchietti2007}.

\begin{figure}[htbp]
    \centering
    \includegraphics[width=0.5\textwidth]{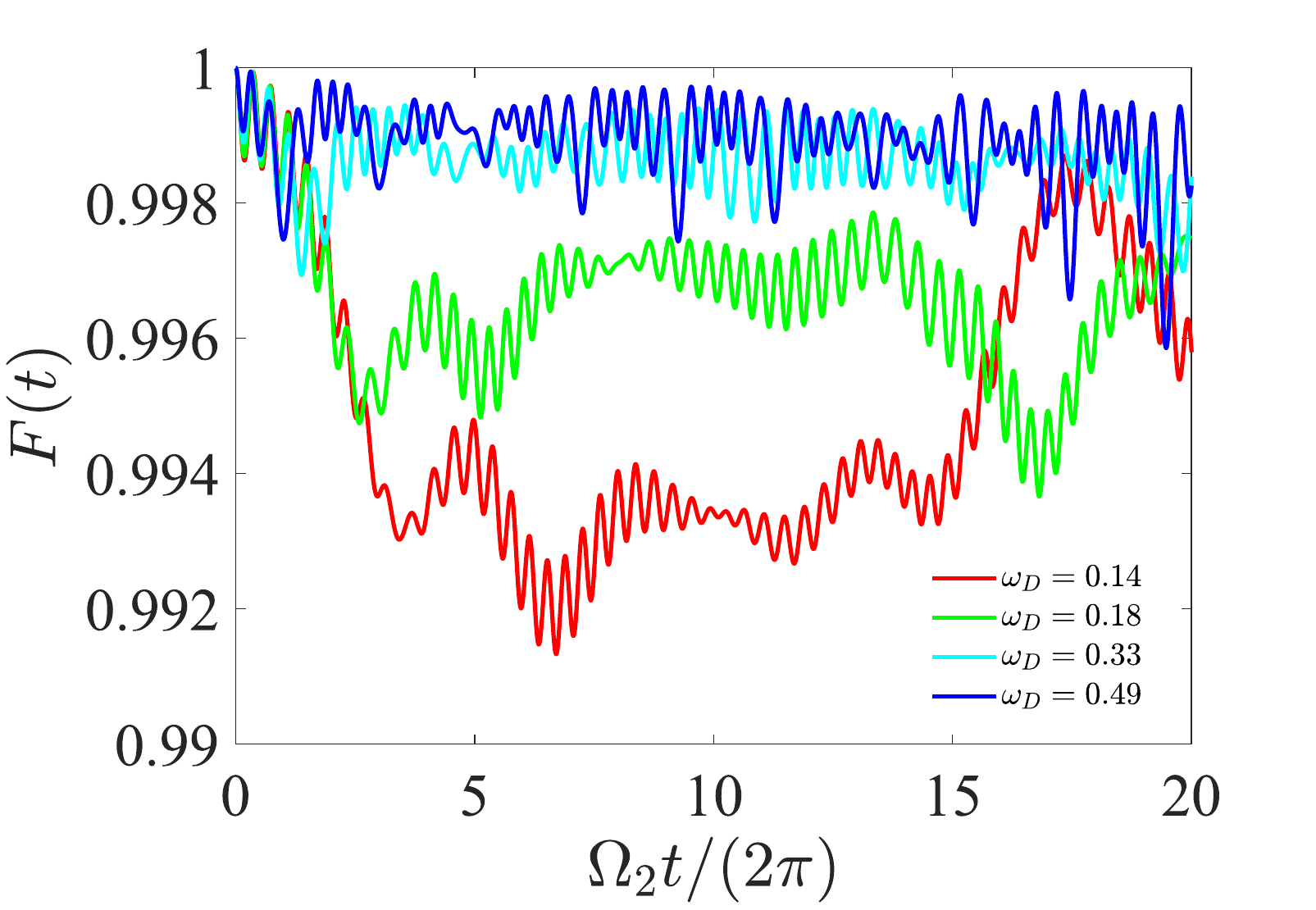} 
    \caption{Loschmidt echo evaluated for different driving frequency values between the states evolved with $H_{rot}$ and $H_{eff}$, respectively, starting from $|\psi_{initial}\rangle$. The values of non-varying parameters are set to $\Omega_{1}=1.25$, $\omega_1=0.5$, $\omega_2=0.25$, $g_1=g_2=0.05$. }
    \label{fig:6}
\end{figure}

\begin{figure}[htbp]
    \centering
    \includegraphics[width=0.5\textwidth]{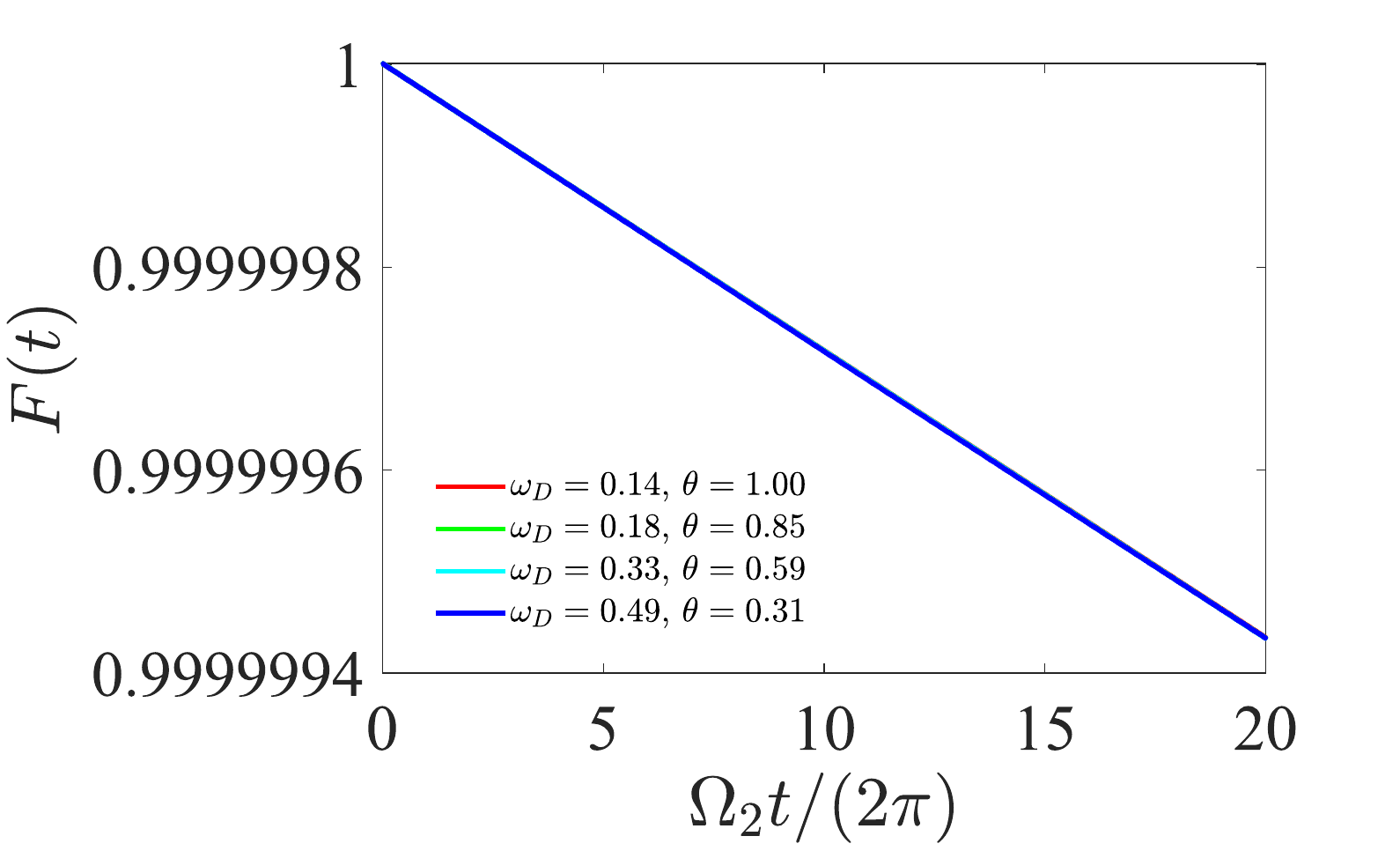} 
    \caption{The plot of time-evolved ground state fidelity obtained between wavefunctions of Hamiltonians $\tilde{H}_{\textit{eff}_{1}}$ and $\tilde{H}_{3JC}$, showing clear agreement between the two states for a significant range of time with $\Omega_1=1.25$. $\omega_1$ and $\omega_2$ takes the same values as Fig.~\ref{fig:6}. }
    \label{fig:7}
\end{figure}

The time evolved states $\ket{\psi_{exact}(t)}$ and $\ket{\psi_{approx}(t)}$ are obtained numerically by solving the time-dependent Schrodinger equation using their respective Hamiltonian starting from the same initial state $\ket{\psi_{initial}}$. This initial state is taken to be $\ket{\psi_{initial}}=\ket{2_{at},\alpha_{1},\alpha_{2}}$ where mode-1 and mode-2 of the cavity are in coherent states denoted by $\ket{\alpha_{1}}$ and $\ket{\alpha_{2}}$ with $\alpha_{1}=\alpha_{2}=0.01$ where its form is given by,
\begin{eqnarray}
    \ket{\alpha_i}&=&\exp(-\frac{1}{2}\abs{\alpha_i}^{2})\sum_{\substack{N_{ph} = 0}}^{\infty}\frac{\alpha_{i}^{N_{ph}}}{\sqrt{N_{ph}!}}\ket{N_{ph}},\nonumber\\
  where  \quad i&=&\{1,2\} \quad and \quad N_{ph}=\{n_{ph},m_{ph}\}
\end{eqnarray}

\noindent with zero detunings, i.e., $\delta_{1}=0$ and $\delta_{2}=0$. To check the consistency of the approximations in Eq.~\ref{eq:11}, we plot in Fig.~\ref{fig:6} the fidelity between $H_{rot}$ and $H_{eff}$ for various driving frequencies $\omega_{D}$. It can be observed that the Loschmidt echo shows an oscillatory behavior between $1$ and $0.991$, indicating that $H_{eff}$ provides a good approximation for $H_{rot}$. Similarly, the plot in Fig.~\ref{fig:7} shows the fidelity between $\tilde{H}_{\textit{eff}_{1}}$ and $\tilde{H}_{3JC}$ with respect to various $\omega_{D}$ and their $\theta$ cutoffs. The values of $\theta$ are chosen such that both $g_{c_{1}}(\theta)$ and $g_{c_{2}}(2\theta)$ remain well below 0.01 simultaneously (see Fig.~\ref{fig:gc}), thereby ensuring that the coupling strengths associated with the counter-rotating terms are negligibly small and can be safely disregarded. This is reflected by the fact that in Fig.~\ref{fig:7} the time evolved fidelities decrease gradually yet does not descend below $0.9$, implying that Eq.~\ref{eq:17} provides a reasonably good approximation for obtaining $\tilde{H}_{3JC}$. Lastly for the sake of confirmation we also verified the behavior of Loschmidt Echo for a set of initial states whose atomic part are given by coherent superposition of two atomic states \cite{Fleischhauer2000}. The other three states are explicitly defined as follows,
\begin{eqnarray}
 \ket{\psi_{initial}}&=& \frac{1}{\sqrt{2}}(\ket{1}_{at}-\ket{2}_{at})\otimes\ket{\alpha_{1}}\otimes\ket{\alpha_{2}},\nonumber\\
 &=& \frac{1}{\sqrt{2}}(\ket{1}_{at}+\ket{3}_{at})\otimes\ket{\alpha_{1}}\otimes\ket{\alpha_{2}},~ \text{and},\nonumber\\
 &=&\frac{1}{\sqrt{2}}(\ket{2}_{at}+\ket{3}_{at})\otimes\ket{\alpha_{1}}\otimes\ket{\alpha_{2}}.  
\end{eqnarray}

\noindent The time-evolved fidelities of all these states were observed to fluctuate between 1 and 0.99, indicating that the approximations introduced in Eq.~\ref{eq:11} and Eq.~\ref{eq:17}, remain highly accurate throughout the evolution. In the next section, we will explore the out-of-equilibrium phase diagram for the driven 3L-JC Hamiltonian.

\section{Phase transitions}
\begin{figure}[htbp]
    \centering
    \includegraphics[width=0.5\textwidth]{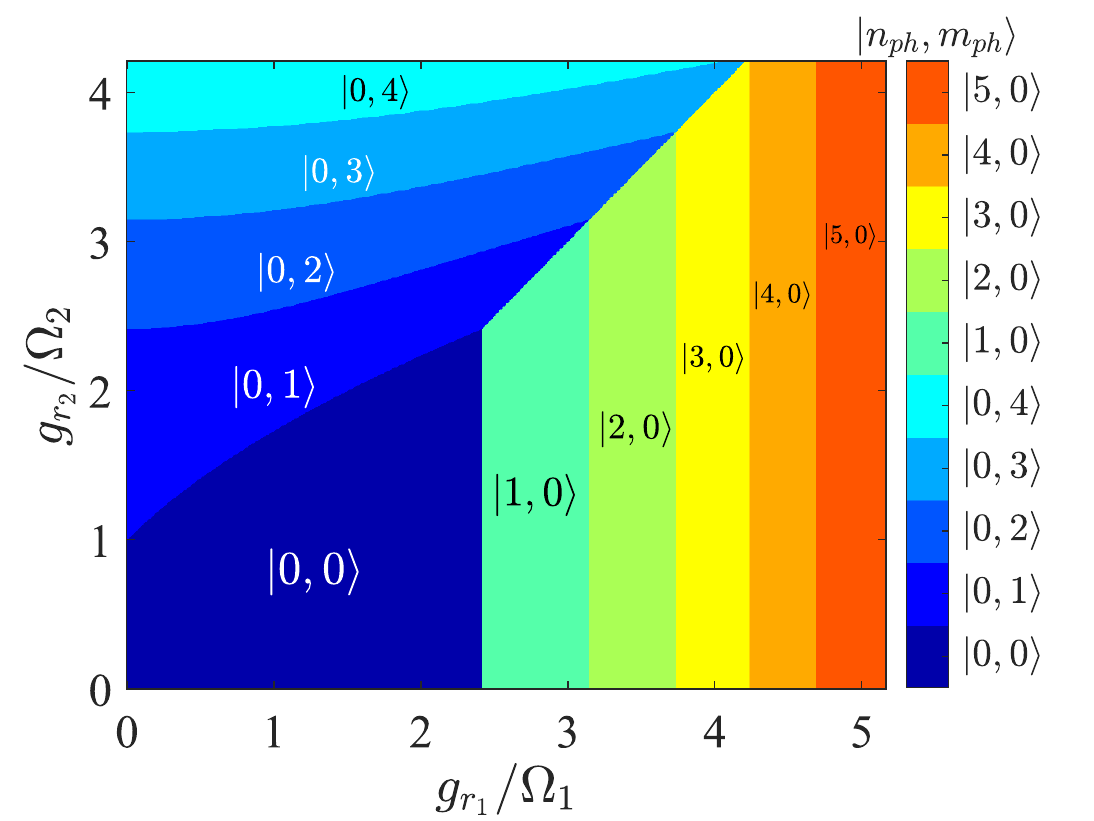} 
    \caption{Plots showing phases arising due to variation of $g_{r_{1}}/\Omega_{1}$ and $g_{r_2}/\Omega_{2}$, for $\delta_{1}=\delta_{2}=0$, $g_{r_{1/2}}\in[0,0.055]$, $\omega_1=0.5$, $\omega_2=0.25$, $\Omega_1=1.25$ and $\Omega_2=1$.}
    \label{fig:example}
\end{figure}

 In this section, we discuss various ground state phases achieved by controlling the driving parameters and validate whether it is possible to access different types of bimodal states $\ket{n_{ph},m_{ph}}$ reported in Fig.~\ref{fig:2}. Initially we begin with examining the phases arising by varying the driving parameters represented by ratio $2\theta=2A_{D}/\omega_{D}$ with respect to the mode-2 detuning $\delta_{2}$ while keeping mode-1 detuning $\delta_{1}$ fixed, since $\ket{3}_{at}\leftrightarrow\ket{2}_{at}$ transition is being modulated.
\begin{figure}[ht]
    \centering
    \begin{subfigure}{0.5\textwidth}
        \centering
        \includegraphics[width=\textwidth]{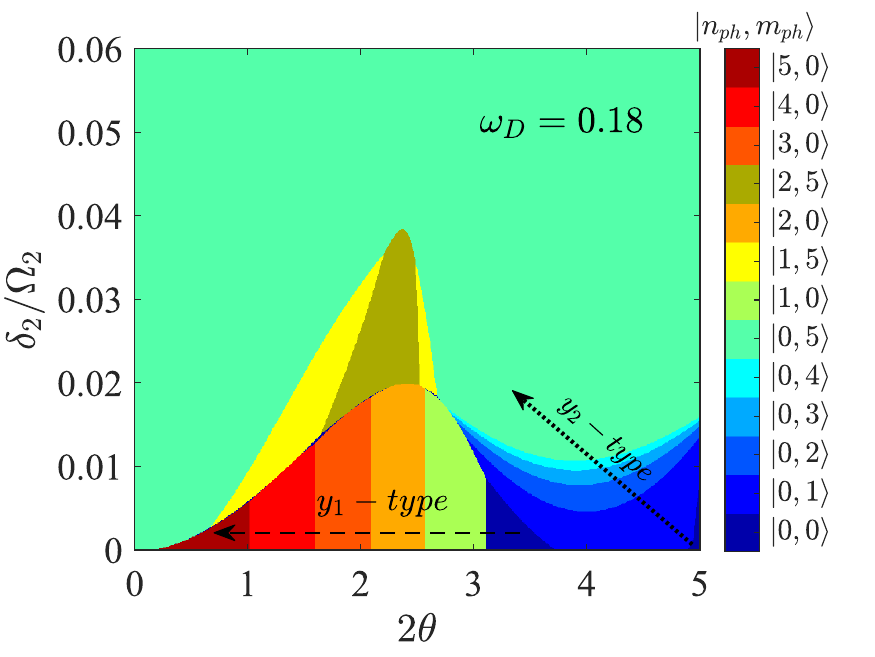}
        \caption{Figure depicting $y_1$, $y_2$ and mixed phases resulting from the variation of detuning $\delta_{2}$ scaled w.r.t mode-2 frequency of the cavity ($\delta_{2}/\Omega_{2}$) and the driving parameters ($2\theta = 2A_{D}/\omega_{D}$) while keeping $\Omega_{1}=1.25$ fixed (i.e., $\delta_1=0$).}
        \label{fig:8(a)}
    \end{subfigure}
    \hfill
    \begin{subfigure}{0.5\textwidth}
        \centering
        \includegraphics[width=\textwidth]{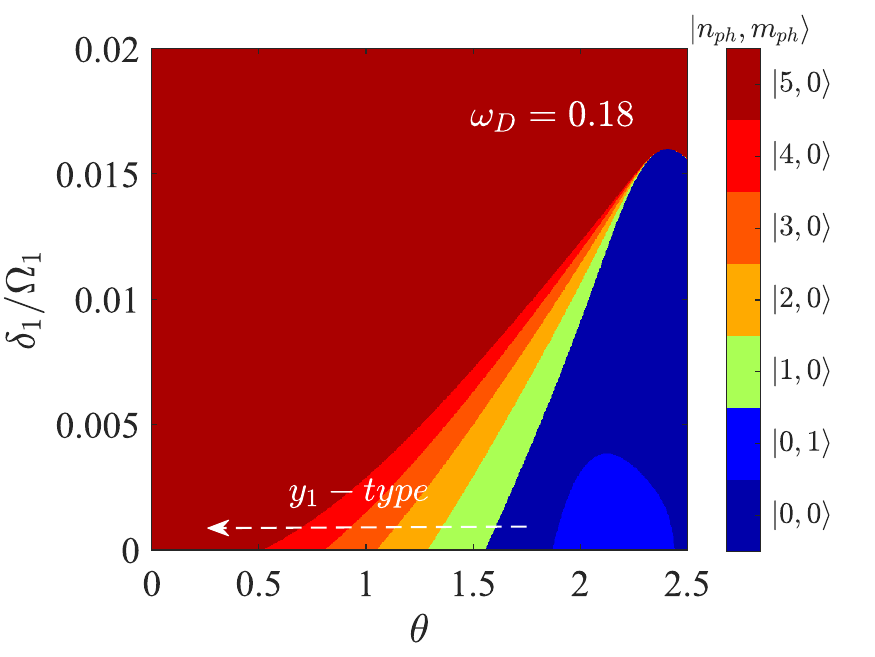}
        \caption{Phase plot achieved by varying the detuning ($\delta_{1}/\Omega_{1}$) and driving parameters ($\theta$). Here $\Omega_{2}$ is set to $\Omega_{2}=1$ which makes $\delta_2=0$.}
        \label{fig:8(b)}
    \end{subfigure}
    \caption{ Plots illustrating distinctive phases obtained by varying the detuning $\delta_{2/1}$ scaled w.r.t frequency of the respective cavity mode  ($\delta_{2/1}/\Omega_{2/1}$) and the driving parameters ($2\theta/\theta$). In both plots $\boldsymbol{(a)}$ and $\boldsymbol{(b)}$ the coupling strengths are taken to be $g_{1}=$ $g_{2}=0.05$ with sidebands $(n_{0},m_{0})=(-14,-11)$ and non-varying atomic energy levels $\omega_1=0.5$, $\omega_2=0.25$. It is to be noted that variation of $\theta$ for a fixed $\omega_{D}$ (here in case $0.18$) comes from varying the driving amplitude $A_{D}$.}
    \label{fig:8}
\end{figure}
 This is shown in Fig.~\ref{fig:8}(\subref{fig:8(a)}) where for a fixed modulating frequency $\omega_{D}=0.18$, we get all the types of phases- $y_1$-type, $y_2$-type as well as mixed-type bimodal states by changing the driving amplitude $A_D$. When we decrease $2\theta$ from $5$ to $3.9$ while simultaneously increasing $\delta_{2}/\Omega_{2}$ from $0$ to $0.015$, along the dotted arrow in Fig.~\ref{fig:8}(\subref{fig:8(a)}),  we observe transition between $y_2$-type phases in the order $\ket{0,0}\rightarrow\ket{0,1}\rightarrow\ket{0,2}\rightarrow\ket{0,3}\rightarrow\ket{0,4}\rightarrow\ket{0,5}$. However, these transitions are expected since $m_{ph}$ represents the photon number in mode-2 coupled to the $\ket{3}_{at}\leftrightarrow\ket{2}_{at}$ transition that is being driven. Nevertheless, we can still access $y_{1}$-type phases by reducing $2\theta$ from $3.4$ to $0.7$ provided that the scaled detuning $\delta_{2}/\Omega_{2}$ remains below $0.003$. As one can observe from Fig.~\ref{fig:8}(\subref{fig:8(a)}), along the dashed arrow the $y_{1}$-type phases evolve in the following sequence, $\ket{0,0}\rightarrow\ket{1,0}\rightarrow\ket{2,0}\rightarrow\ket{3,0}\rightarrow\ket{4,0}\rightarrow\ket{5,0}$. {In the same figure we also notice two prominent mixed type phases $\ket{1,5}$ and $\ket{2,5}$ arising near the center. Although these are not the only mixed phases present in the phase diagram, the other mixed phases are too small to be resolved at the chosen scales of $\delta_2/\Omega_2$ and $2\theta$.} One crucial thing to observe is that when $2\theta$ is negligibly small we get a higher order phase $\ket{0,5}$ over a wide range of $\delta_{2}$ even though we have kept $g_{1}=g_{2}=0.05$, but in the static Hamiltonian (Eq.~\ref{eq:2}) if we set $g_{1}=g_{2}=0.05$, one would only observe a normal-phase $\ket{0,0}$ as clearly shown in Fig.~\ref{fig:2}. However, owing to the external driving, the coupling terms $g_1,\; g_2$ are now replaced by the effective couplings $g_{r_1}$ and $g_{r_2}$ respectively, which are much larger than the critical coupling values confining the normal phase of the undriven Hamiltonian. Therefore, it is expected to observe a higher-order state of the static Hamiltonian as the ground state of the driven Hamiltonian. Similarly in Fig.~\ref{fig:8}(\subref{fig:8(b)}) we see phases varying from $\ket{0,0}\rightarrow\ket{1,0}\rightarrow\ket{2,0}\rightarrow\ket{3,0}\rightarrow\ket{4,0}\rightarrow\ket{5,0}$ (along the white dashed arrow) containing only $y_1$-type phases. Meanwhile, we also get a dome-shaped $\ket{0,1}$ phase which is of $y_2$-type inside the normal phase region $\ket{0,0}$ for a larger value of $\theta$. These variation in phases in Fig.~\ref{fig:8}(\subref{fig:8(b)}) even though $\ket{3}_{at}\leftrightarrow\ket{1}_{at}$ transition is not driven arises due to the fact that both $\ket{1}_{at}$ and $\ket{2}_{at}$ are coupled to common energy level $\ket{3}_{at}$. Similar to the phase diagram in Fig.~\ref{fig:8}(\subref{fig:8(a)}) the immediate occurrence of higher order phase $\ket{5,0}$ at negligible values of $\theta$ can be explained by the  Fig.~\ref{fig:gr}(c), where it can be observed that the effective coupling strengths increases beyond the critical coupling value (of the static Hamiltonian) as $\theta$ approaches zero. It is to be noted that, moving along the $y_1$-type dashed arrow line in Fig.~\ref{fig:8}(\subref{fig:8(a)}) and Fig.~\ref{fig:8}(\subref{fig:8(b)}), for the limiting case of both $\delta_{2}/\Omega_{2}$ and $\delta_{1}/\Omega_{1}$ approaching zero, identical phase sequence appears in the order $\ket{0,0}\rightarrow\ket{0,1}\rightarrow\ket{0,0}\rightarrow\ket{1,0}\rightarrow\ket{2,0}\rightarrow\ket{3,0}\rightarrow\ket{4,0}\rightarrow\ket{5,0}$, as $\theta$ is decreased from the far right values in plot up-till the head of the arrow. Finally, in Fig.~\ref{fig:example} we show phases arising from variations of effective coupling parameters $g_{r_{1}}$ and $g_{r_{2}}$, both normalized to their respective cavity mode frequencies of the static Hamiltonian. We observe a ground state phase diagram similar to that in Fig.~\ref{fig:2} but with two orders of magnitude smaller than the undriven coupling values $g_1/\Omega_1$ and $g_2/\Omega_2$.  We normalized the coupling strengths with respect to cavity frequencies ($\Omega_{1}$, $\Omega_{2}$) of the static Hamiltonian instead of effective cavity frequencies ($\tilde{\Omega}_{1}$, $\tilde{\Omega}_{2}$) to demonstrate that even when the $g_1$ and $g_2$ are taken to be small values (both $\in[0,0.055]$ for Fig.~\ref{fig:example}), it is possible to observe higher order phases as the values of effective couplings are changed in the same manner as in Fig.~\ref{fig:2}. All of above phase diagrams in this section have been obtained by diagonalizing the effective 3L-JC Hamiltonian $\tilde{H}_{3JC}$, and carefully sorting their ground state energies in order to finally compare them with the eigen-energies phase-diagram in Fig.~\ref{fig:2}.

 As for the experimental implementation of such system, they can be realized in a single InAs/GaAs quantum dot which under the application of an external magnetic field along direction perpendicular to the growth axis, exhibits a double $\Lambda$-type scheme \cite{Burgers, Schaibley2013}. Using suitable choice of two orthogonally polarized cavity modes (eg.:- $\Omega_{1}\rightarrow H$ polarized and $\Omega_{2}\rightarrow V$ polarized) obtained in a planar microcavity structure one can reduce the double $\Lambda$-type scheme into a single $\Lambda$-type 3LS interacting with two cavity modes. The corresponding mapping of states are given by $\ket{x^{+}}=\ket{1}_{at}$, $\ket{x^{-}}=\ket{2}_{at}$ and $\ket{T_{x}^{-}}=\ket{3}_{at}$ where $\ket{x^{+}}$, $\ket{x^{-}}$ are electron spin states while $\ket{T_{x}^{-}}$ is negative trion state. This mapping ensures that the experimental setup is reasonably consistent with our model. 



\section{Conclusion}
In summary, we presented various bimodal superradiant phases for a $\Lambda$-type 3LS embedded in a lossless double-mode cavity. The critical coupling strengths between the normal and the superradiant phases are marked by the $\ket{0,0}$ phase, shown in Fig.~\ref{fig:2}. It is beyond this region that the system exhibits a bimodal superradiant phase transition which trifurcates into three distinct categories, $y_1$, $y_2$, and mixed-type phases, depending on the values of $g_1/\Omega_1$ and $g_2/\Omega_2$. Further, we propose a method to show that these superradiant phases can be achieved for the system under consideration without compromising the RWA, i.e., without exceeding the critical coupling limits defined by the normal-phase $\ket{0,0}$ region. This is accomplished by periodically modulating only a subspace of $\Lambda$-type system (i.e., only the $\ket{3}_{at}\leftrightarrow\ket{2}_{at}$ transition) Eq.~\ref{eq:4} and performing a series of unitary transformations (Eq.~\ref{eqn:5},.~\ref{eq:u2}) to obtain the effective 3L-JC Hamiltonian. It was observed that this effective Hamiltonian can now support higher-order superradiant phases even for a small value of $g_1$ and $g_2$ because of the periodic modulation. In comparison with the periodic modulation of a two-level system (2LS) as studied in Ref.~\cite{ChengLiu2023}, we note that the presence of a single cavity mode in their model leads to the emergence of the normal phase over a broad region in the detuning–driving amplitude parameter space. In contrast, our bimodal cavity system exhibits a significantly larger domain where higher-order phases appear, highlighting the enhanced complexity and richness of the phase structure in the presence of two interacting modes. Our results may facilitate others to explore phase transitions in finite-sized quantum optical models beyond 2LS.


\appendix

\section{Hamiltonian in bare state eigenbasis} \label{Appndx:Appendix A}

We assume the bare states of the undriven Hamiltonian $H_{JC}$ (as shown in Eq.\ref{eq:2}) are given by :- 


\begin{align}
\ket{\psi^{1}} & = \ket{3}_{at} \ket{n_{ph}} \ket{m_{ph}-1},\nonumber\\
    \ket{\psi^{2}} & = \ket{1}_{at} \ket{n_{ph}+1} \ket{m_{ph}-1}, \nonumber \\
    \ket{\psi^{3}} & = \ket{2}_{at} \ket{n_{ph}} \ket{m_{ph}},    
\end{align}

\noindent where $\ket{1}_{at}$, $\ket{2}_{at}$, $\ket{3}_{at}$ are atomic energy level states mapped to bare states $\ket{\psi^{2}}$,$\ket{\psi^{3}}$ and $\ket{\psi^{1}}$ while the later two are photonic states in Fock state basis. Writing $H_{JC}$ in the eigenbasis of the above three bare states we get the following Hamiltonian,

\begin{eqnarray}
    H_{JC} = \begin{pmatrix}
  H_{11} & H_{12} & H_{13} \\
  H_{21} & H_{22} & H_{23}  \\
  H_{31} & H_{32} & H_{33}          
\end{pmatrix}\label{hnm}
\end{eqnarray}

\noindent with elements given by,

\begin{align} \label{eq:A2}
    H_{11}= \bra{\psi^{1}}H\ket{\psi^{1}}&= \omega_{1}+\omega_{2}+\Omega_{1}n_{ph}+\Omega_{2}(m_{ph}-1),  \nonumber \\
    H_{12}= \bra{\psi^{1}}H\ket{\psi^{2}}&=g_{1}\sqrt{n_{ph}+1},  \nonumber \\
    H_{13}= \bra{\psi^{1}}H\ket{\psi^{3}}&=g_{2}\sqrt{m_{ph}},  \nonumber \\
    H_{21}= \bra{\psi^{2}}H\ket{\psi^{1}} & = g_{1}\sqrt{n_{ph}+1},  \nonumber \\
    H_{22}= \bra{\psi^{2}}H\ket{\psi^{2}} & =-\omega_{1}+\Omega_{1}(n_{ph}+1)+\Omega_{2}(m_{ph}-1),  \nonumber \\
    H_{23}= \bra{\psi^{2}}H\ket{\psi^{3}} & = 0,\nonumber \\
    H_{31}=\bra{\psi^{3}}H\ket{\psi^{1}}&=g_{2}\sqrt{m_{ph}},  \nonumber \\
    H_{32}= \bra{\psi^{3}}H\ket{\psi^{2}} & = 0 ,  \nonumber \\
   H_{33}= \bra{\psi^{3}}H\ket{\psi^{3}}&=-\omega_{2}+\Omega_{1}n_{ph}+\Omega_{2}m_{ph}.
\end{align}

\noindent {As explained in detail in the main text, $\omega_1$, $\omega_2$, $\Omega_1$, and $\Omega_2$ are the atomic and cavity mode frequencies for the three-level JC Hamiltonian, with $g_1$ and $g_2$ being coupling constants. This $3\times3$ Hermitian Hamiltonian has real eigenvalues and orthonormal eigenvectors. However, analytically, the eigenvalues are roots of a cubic polynomial; therefore, the eigenvectors and eigenvalues are found by solving the
eigenvalue equation numerically for practical purposes. Consequently, we have used MATLAB to extract the numerical solution of the above Hamiltonian in order to plot the ground-state phase diagram represented in Fig.~\ref{fig:2}.Whereas, for the final effective Hamiltonian in Eq.~\ref{eq:H3JC}, we will replace all the parameters in Eq.~\ref{eq:A2} by their effective values i.e., $\tilde\omega_{1/2},~\tilde{\Omega}_{1/2}$, and $g_{r_{1/2}}$.}


\bibliography{Main_ThreelevelLambdaJC}
\end{document}